\documentclass[letterpaper, 10 pt, conference]{ieeeconf}
\IEEEoverridecommandlockouts
\usepackage{cite}
\usepackage{amsmath,amssymb,amsfonts,bm}
\usepackage{algorithmic}
\usepackage{graphicx}
\usepackage{textcomp}
\usepackage{url}
\usepackage{bbding}
\usepackage{listings}%
\usepackage{hyperref}
\usepackage{subfigure}
\usepackage{diagbox}
\usepackage{makecell}
\usepackage{float}
\usepackage[ruled,vlined,linesnumbered]{algorithm2e}
\SetKwComment{Comment}{$\triangleright$\ }{}
\usepackage{booktabs, tabularx}
\newcolumntype{C}{>{\centering\arraybackslash}X}   

\title{\LARGE \bf
Multi-FEAT: Multi-Feature Edge Alignment \\ for Targetless Camera-LiDAR Calibration
}

\author{Bichi Zhang$^{1}$, Holger Caesar$^{2}$ and Raj Thilak Rajan$^{1, ^\dagger}$ 
\thanks{*This work is partially funded by the European Leadership Joint Undertaking
(ECSEL JU), under grant agreement No. 876019, the ADACORSA project \cite{ADACORSA}.}%
\thanks{$^1$ Faculty of Electrical Engineering, Mathematics and Computer Science, TU Delft, the Netherlands.}%
\thanks{$^2$ Faculty of Mechanical Engineering, TU Delft, the Netherlands.}%
\thanks{$^\dagger$ Corresponding author.}%
}

\begin{document}


\maketitle
\thispagestyle{empty}
\pagestyle{empty}

\begin{abstract}
Multi-agent systems, e.g., automobiles and UAVs (Unmanned Ariel Vehicles), rely on the precision of onboard sensors to accurately perceive their environment, which in turn depends on the precision of onboard sensors and reliable in-field calibration. This paper introduces a novel targetless camera-LiDAR extrinsic calibration approach called Multi-FEAT (Multi-Feature Edge AlignmenT). Multi-FEAT uses the cylindrical projection model to encode the 3D LiDAR point cloud into a 2D panorama and exploits diverse LiDAR feature information in panoramic images to supplement the sparse LiDAR point cloud boundaries. Furthermore, camera edges are extracted using off-the-shelf segmentation solutions. In addition, a feature-matching function is designed to optimize the calibration parameters. The performance of the proposed Multi-FEAT algorithm is evaluated using the KITTI dataset, and our approach shows more reliable results than several existing targetless calibration methods. We conclude our analysis with directions for future work.
\end{abstract}



\section{Introduction}
Multi-sensor fusion techniques are conventionally employed to enhance the environmental perception and localization capabilities of autonomous agents. However, a critical challenge in achieving effective multi-sensor fusion is the calibration of multiple sensors, which involves determining the spatial relationship between sensors through rotation and translation parameters. The primary task of sensor calibration is to establish data associations between sensors, thereby enabling accurate estimation of their relative poses.
This task becomes particularly challenging when calibrating sensors of different modalities, such as in the case of camera-LiDAR calibration. LiDAR sensors provide accurate but sparse 3D positional information of the surroundings, while cameras capture rich visual signals for scene understanding but lack distance information. Although camera-LiDAR fusion offers complementary characteristics, it poses difficulties in data association during calibration tasks.

Traditionally, calibration methods rely on known objects in the field, such as checkerboards, to align common features across different sensor modalities \cite{zhangqilong}. While target-based methods have historically provided reliable solutions, their reliance on human intervention is a significant drawback, particularly in automotive systems, prompting the development of targetless calibration methods. Some targetless calibration approaches utilize direct measurements from both sensors, such as laser reflectivity and image intensity, and associate them by maximizing specific target functions, such as mutual information or cross-correlation \cite{MI}. Another category of methods leverages geometric information encoded in LiDAR point clouds, such as depth discontinuities \cite{Levinson} and surface norms \cite{GOM}, to align with image intensity edges. To compensate for the lack of depth information in camera images, some methods adopt structure-from-motion techniques \cite{spatio_temporal} \cite{sen2023scenecalib} or employ stereo \cite{hu2022tescalib} or multi-camera \cite{multi_camera_lidar} setups to associate geometric information from cameras with LiDAR measurements. However, motion-based methods require motions of the platform, which could be fatally dangerous for autonomous vehicles maneuvering with uncalibrated sensors. The introduction of stereo camera pairs does provide additional information, but at a cost of a heavier computational burden.

\begin{figure}[t]
\centering
\includegraphics[width=0.45\textwidth]{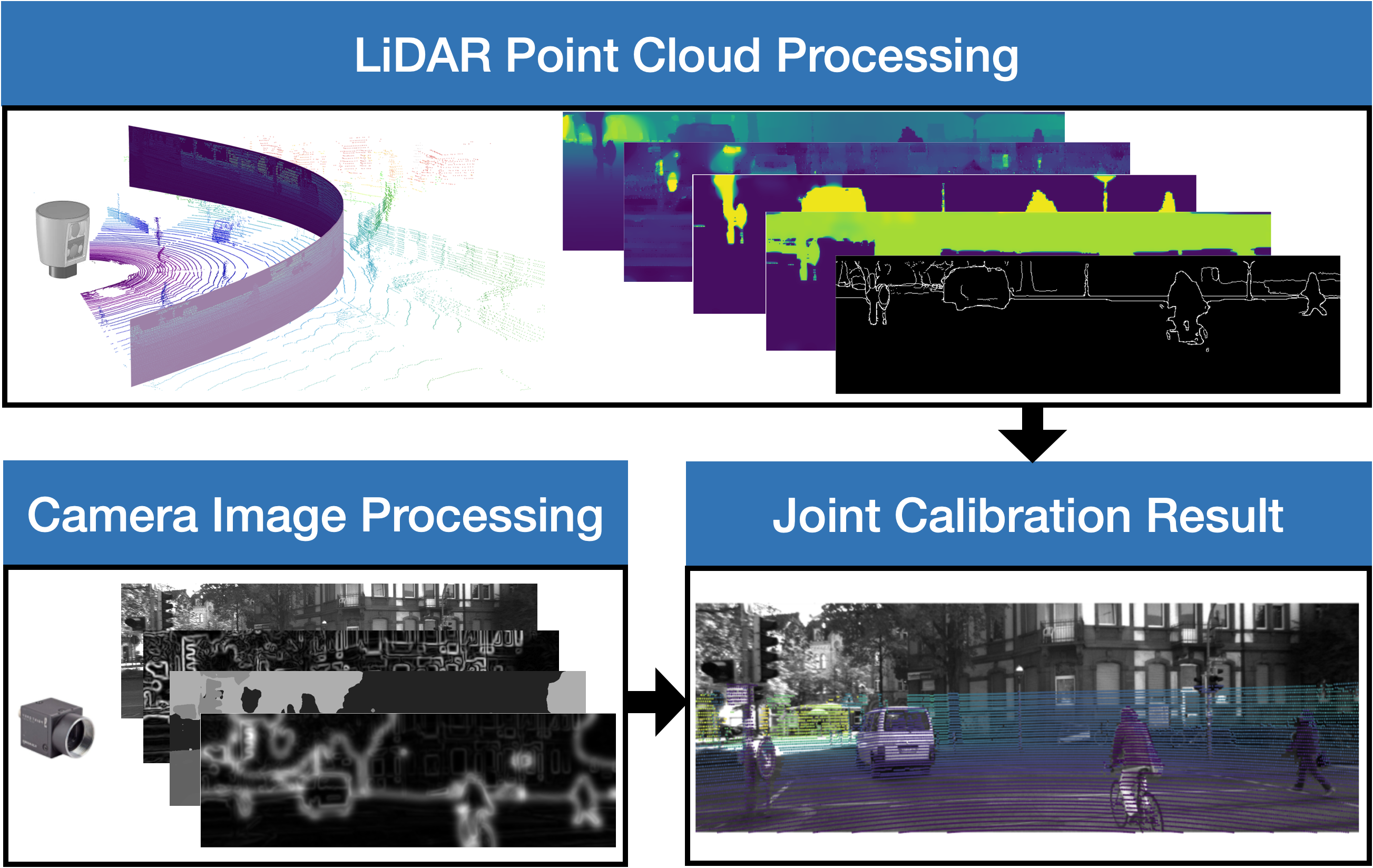}
\caption{\small A visual overview of the proposed Multi-FEAT algorithm.}
\label{fig:main}
\end{figure}

In this work, we propose our targetless camera-LiDAR calibration pipeline Multi-FEAT: Multi-Feature Edge AlignmenT, which is briefly illustrated in Fig \ref{fig:main}. Multi-FEAT leverages multiple features in the field of view (FOV) of LiDAR for better utilization of the laser information and generates an enriched multi-feature edge map through the extraction of various features from the LiDAR point cloud following a cylindrical projection. We adopt Grounded Segment Anything Model (Grounded-SAM), a state-of-the-art pipeline that feeds the Segment Anything Model (SAM) \cite{kirillov2023segment} with object detection results from Grounding DINO \cite{liu2023grounding} as prompts, to create instance-level semantic edges from the input images, and enhance them with intensity information. The calibration parameters are calculated via optimization of the designed edge alignment target function, which acquires both processed camera and LiDAR edge maps. In summary, our key contributions to this work are as follows:
\begin{itemize}
    \item We propose a targetless calibration method called Multi-FEAT, which solves the camera-LiDAR extrinsic calibration problem by exploiting multiple features from point clouds and camera images.
    \item We convert the 3D-2D association problem into a 2D-2D registration problem, and design a cost function to align the edge intensities from both the camera edge map and the multi-feature LiDAR edge map.
    \item We compare our proposed Multi-FEAT algorithm with the state-of-the-art targetless calibration methods, using the open-source KITTI dataset \cite{KITTI_2012}, and show that our proposed solution outperforms the existing methods even in harsh environments. 
\end{itemize}

\subsection{Related Works}\label{sec:related_works}
In essence, the joint camera-LiDAR calibration task involves estimating the relative pose between these two sensors. Conventionally, this challenge is addressed by using known external reference objects in the field. For instance, single checkerboards \cite{zhangqilong} \cite{Unnikrishnan} and multiple checkerboards placed at various locations \cite{Geiger} have been commonly employed. In recent years, there has been a growing exploration of alternative design objects with varying shapes, such as rings \cite{ring_2008}, a trihedron \cite{trihedron_2013}, a sphere \cite{target_ball_2016}, V-shape objects \cite{vshape_2016} \cite{multiboard2_2018}, and cuboids \cite{cuboid_2017}.
However, the reliance on artificial objects imposes limitations on the feasibility and restricts the applicability of these solutions to controlled environments like laboratories and test fields. The development of a targetless or object-less calibration solution holds significant promise as it not only reduces the need for human intervention but also paves the way for commercial applications, particularly in autonomous vehicular systems \cite{Yeong_overview}. Furthermore, this approach holds potential for futuristic autonomous space systems \cite{bajracharya2008}.

The establishment of direct data associations between different sensor modalities becomes particularly challenging in the absence of artificial targets. Targetless methods seek to address this challenge via statistical correlations between different sensory measurements. An early approach borrowed from image registration for targetless camera-LiDAR calibration is to maximize the mutual information between the grayscale intensities of the camera image and the reflectivity values of the LiDAR \cite{MI}. An alternative to reflectivity is to use the depth discontinuity of the LiDAR output to capture the edge information of any object that is ubiquitous in the road environment \cite{Levinson} \cite{Irie}, or to use a gradient orientation measurement to match the LiDAR reflectivity map with the camera grayscale image \cite{GOM}. Another algorithm based on edge matching is the idea of solving the calibration-fusion joint problem \cite{castorena}, where the point cloud is projected onto the camera image plane and the projected sparse points are used for depth completion and match the edges of the depth map and the intensity image using cosine similarity. 

Some attempts of end-to-end networks are also proposed for calibration, e.g., RegNet \cite{Regnet} and CalibNet \cite{calibnet}, which require known ground truth and large-scale datasets for training. In application, when the amount of onboard sensors grows, the deployment cost and generality of such end-to-end networks are challenged. Some other approaches adopt neural networks only for feature extraction. E.g., PSP-Net for vehicle extraction \cite{semantic}, and BiSeNet-v2 \cite{crlf} for pole extraction have been applied for feature matching tasks. However, these methods only include designated semantic features, which are less reliable when the specific objects do not exist in the field. In our work, Multi-FEAT applies an off-the-shelf generalized semantic segmentation network Grounded-SAM \cite{GSAM} and uses the semantic edges as the primary image edge. By exploiting multiple natural features from LiDAR point clouds, Multi-FEAT supplements the shortcomings of several current edge-alignment methods and improves the reliability of the results.

\subsection{Problem Formulation}\label{sec:problem_formulation} 

A point cloud is a set of measurements from LiDAR sensors, denoted as $\bm \psi$, and all the points that belong to the set of the point cloud are given by $\{\mathbf p = [x,y,z] \in \mathbb{R}^3 |\ \mathbf p \in \bm \psi\}$. Each pixel in the camera image is a 2D vector $\mathbf u = [i, j]^T \in \mathbb{R}^2$. Typically, the camera image and the LiDAR point cloud are geometrically related using the following rigid body transformation,
\begin{equation}
\left[\begin{array}{l}
{\mathbf u} \\
1
\end{array}\right]
 = \bm K
\overbrace{\left[\begin{array}{cc}
\bm{R}  & \bm{t}\\
\bm{0} & 1
\end{array}\right]}^{\bm T_{\bm \theta}}
\left[\begin{array}{l}
{\mathbf p}\\
1
\end{array}\right]
\label{eq:extrinsic}
\end{equation} where the camera intrinsic matrix $\bm K$ projects the points from 3D to 2D space and thus forming an image on the 2D plane (\ref{eq:extrinsic}). The matrix $\bm T_{\bm \theta}$ determined by $\bm R$ and $\bm t$ contains the extrinsic calibration parameters, representing the rigid transformation from LiDAR coordinate system to the camera coordinate system and this transformation matrix belongs to the $SE(3)$ Lie group. Here,  $\bm R$ is known as 3D rotation matrix, determined by the Euler angles $r_x, r_y,$ and $r_z$. The translation vector $\bm t$ includes the 3D translational directions, $t_x$, $t_y$ and $t_z$. The goal of edge alignment is to match the edges of the point cloud and the image, subsequently estimate the unknown extrinsic calibration parameters $\bm \theta= [r_x, r_y, r_z, t_x, t_y, t_z]^T$. 

\begin{figure*}[t]
    \centering
    \subfigure[Semantic edge map]{
    \includegraphics[width=0.44\textwidth]{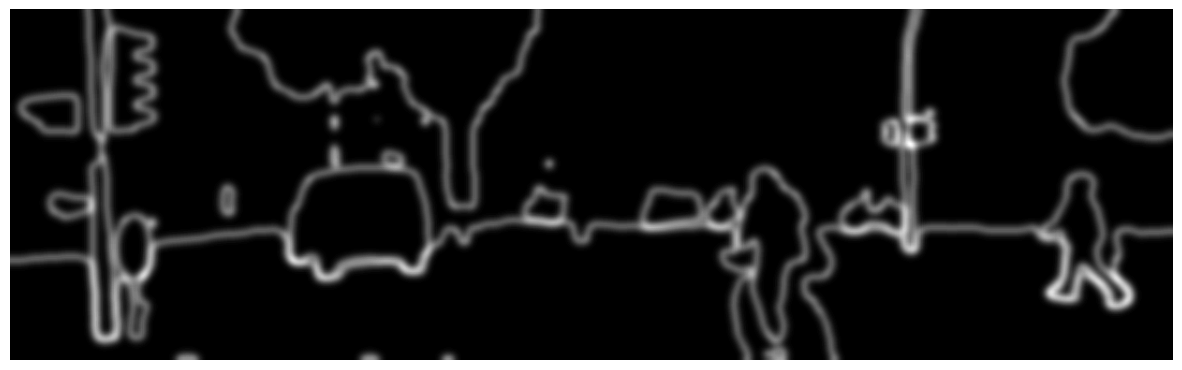}}
    \subfigure[Fused image edge map]{
    \includegraphics[width=0.44\textwidth]{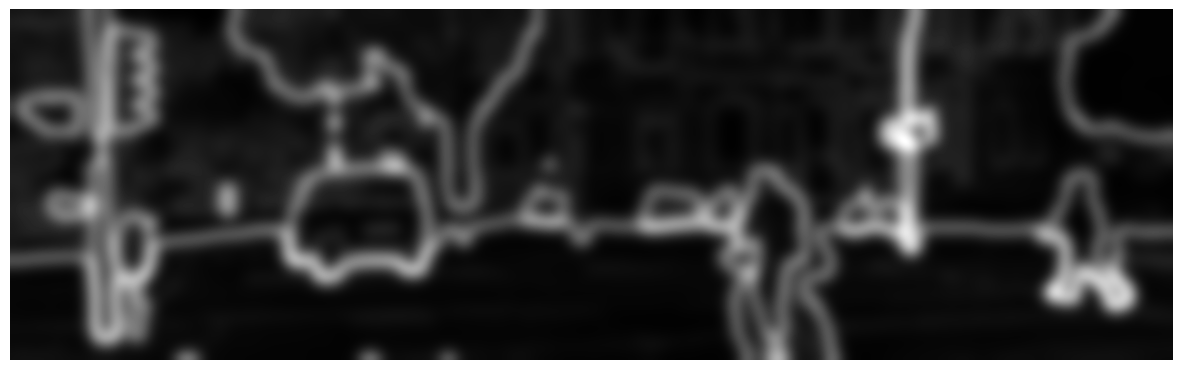}}
    \caption{\small An illustration of image edge fusion: The general edge information can be extracted by the semantic edge extractor, shown in (a). The fused edge map with a low-weight Sobel map compensates the details is shown in (b).}
    \label{fig:img_fusion}
\end{figure*}
\section{Image Processing} \label{sec:im_processing}
For sensor fusion purposes, camera and LiDAR systems usually have a considerable proportion of the overlapping area. However, one crucial concern is that due to different modalities of sensors, cameras and LiDARs do not necessarily observe the same features.
In this section, we focus on extracting edge features from camera images that match with LiDAR point clouds. 

Structural information can be easily extracted from LiDAR measurements, while images provide more visual contents like intensity and luminance.
Conventional methods rely on intensity-based edge extraction, which the light condition could heavily influence. For this problem, the semantic boundaries for isolated objects are much more robust than the intensity edges. Therefore, we adopt an off-the-shelf pre-trained segmentation network SAM \cite{kirillov2023segment} as the primary edge extractor to overcome such influences. However, the granularity of the network is difficult to be set adaptive in different scenarios. Over-segmentation and under-segmentation bring noises which are hard to get rid of. Therefore, we pre-define a dictionary as a priori information, including several commonly seen objects in traffic scenes. Prompted by the bounding boxes identified via Grounding DINO \cite{liu2023grounding}, SAM performs more effectively and efficiently with high quality edges of objects. 
 
Previous research shows that the intensity feature of the image, on the other hand, presents valid correlation between LiDAR reflectivity \cite{MI}. To further enrich the detailed information, we aggregate the Sobel edge map on the semantic edge map with a lower weight. Therefore, the fused edge map promotes the pattern information omitted by the semantic layer while preserving the prominent edges between different semantic areas.
To penalize the points that locate far from the edges, the current edges should be expanded wider to cover larger areas with decaying weight. In \cite{Levinson}, the authors proposed an inverse distance kernel to reward pixels that locate close to edges. We achieve the similar goal by applying a Gaussian kernel on the edge map $\bm E$ to expand the edges and smooth transitions. The term $\bm E_C(\bm u)$, indicates the mapping from a given position $\bm u$ on the image to the edge intensity value.

\section{Point Cloud Processing} \label{sec:PC_processing} Extracting point cloud boundaries is non-trivial since the points are arbitrarily scattered in 3D space without orderly values. In Multi-FEAT, we encode 3D sparse points into a 2D sparse image while preserving the spatial structure using cylindrical projection. However, the LiDAR coordinate plane $X-Y$ is not parallel to the emission direction of the beam. As a result, some foreground points may occluded with the background points, which means after the projection, their pixels on the 2D panorama might be blended. Such effect is of particular severe around the boundaries of foreground objects. To acquire high-quality edge information from the 2D map, we first apply morphological closing to resolve the occlusion effect (\ref{sec:cylindrical projection}) by defining the foreground objects beforehand via clustering. We further enrich the edge information by incorporating depth, reflectivity, foreground objects and the ground plane, generating high-resolution panoramic maps (\ref{sec:dense_map_comp}). Finally, the Canny edge detector is used to extract edge information of the 2D maps, combining them as the multi-feature edge map of the point cloud (\ref{sec:pc_edge}).

\subsection{Foreground Objection Classification} \label{sec:dbscan}
\begin{figure*}[t]
\centering
\subfigure[$\bm{\psi}$]{
\includegraphics[width=0.24\textwidth]{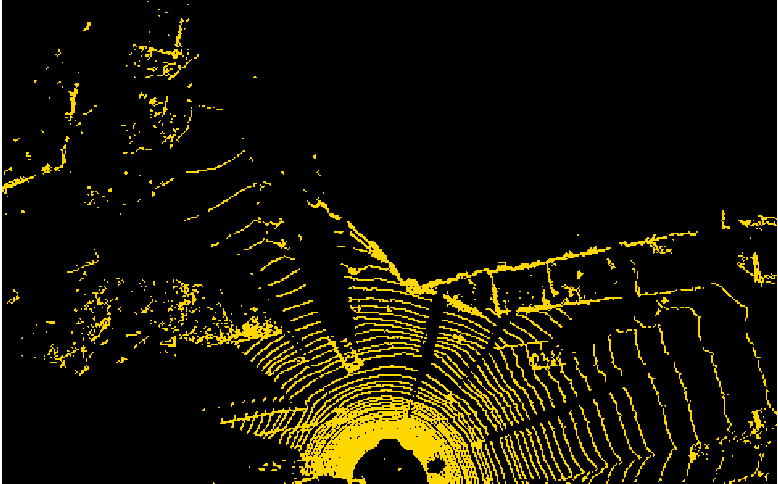}}
\subfigure[$\bm{\psi_1}$]{
\includegraphics[width=0.24\textwidth]{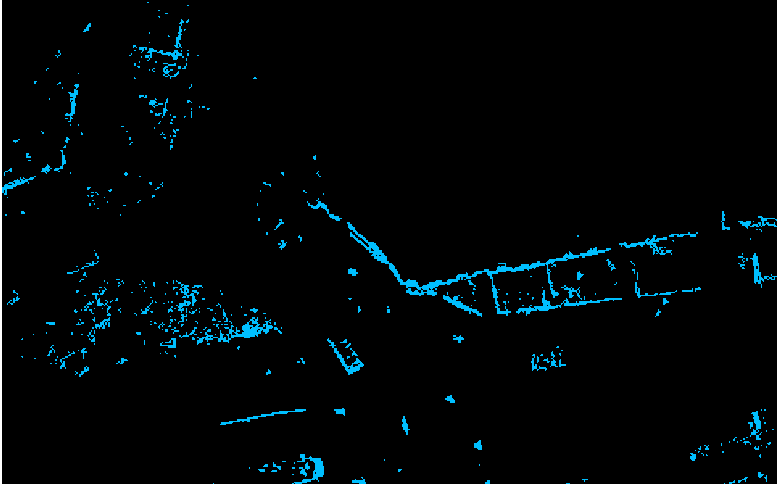}}
\subfigure[$\bm{\psi}$ with labels]{
\includegraphics[width=0.24\textwidth]{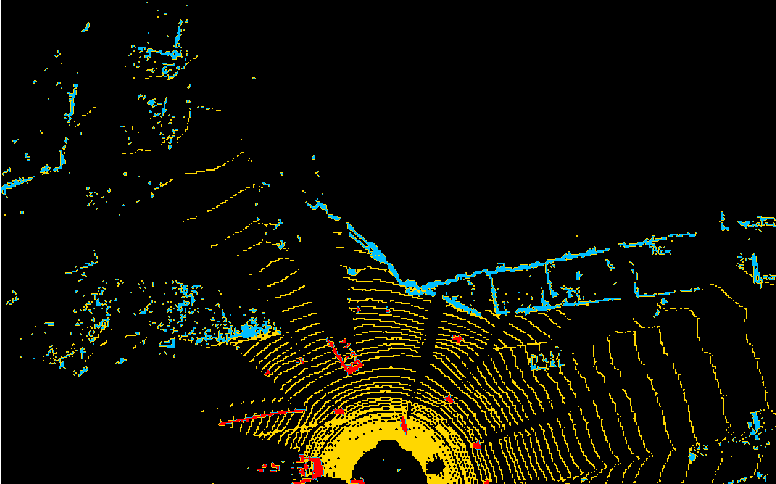}}
\caption{\small Effect of RANSAC and DBSCAN algorithms: (a) The BEV of the raw LiDAR point cloud ($\bm{\psi}$). (b) The processed point cloud ($\bm{\psi_1}$) after applying RANSAC plane segmentation on $\bm{\psi}$ (c) The BEV of the selected foreground objects ($\bm{\psi_2}$) after DBSCAN clustering on $\bm{\psi_1}$ (\ref{sec:dbscan}). The red pixels in the third image represent the foreground objects ($o=2$), while the blue pixels ($o=1$) are back ground and the yellow ones are plane points and outliers ($o=0$).}
\label{fig:RANSAC_DBSCAN}
\end{figure*}

The 3D point cloud $\bm{\psi}$ obtained from the LiDAR captures surroundings comprising foreground and background information. The key feature of point clouds stems from the geometry edges. The foreground objects receive more measures in one frame, therefore preserves higher qualities. However, the edges of foreground objects suffer from the occlusion effect due to non-parallel directions between the projection and the emission direction \cite{premebida2016high}, which decreases the overall quality.
Furthermore, the projected panoramic map also suffers from spurious points which do not belong to an object of interest, e.g., ground points, visualized as arcs in the Birds' eye view (BEV) of the raw point cloud in Figure \ref{fig:RANSAC_DBSCAN}(a). 
We adopt RANSAC to eliminate the interference from ground points and subsequently improve the identifiability of objects in the FOV \cite{RANSAC}. Given the processed LiDAR point cloud $\bm{\psi_1}$ (see Figure \ref{fig:RANSAC_DBSCAN}(b)), we apply DBSCAN to identify densely connected regions. The average depth of a cluster $k$ is marked as $d_k$. Let $o_k \in \mathbb{N}$ denote the flag for the $k$th cluster, indicating if the cluster is part of the near field, background or empty, based on a threshold $\delta$, i.e., \begin{equation} o_k = 
    \begin{cases}
        2, \text{if}\ d_k < \delta\ \text{(foreground)} \\
        1, \text{if}\ d_k > \delta\ \text{(background)} \\
        0, \text{otherwise}\ \text{(empty)}
    \end{cases}
    \label{eq:obj_values}
\end{equation}

\subsection{Cylindrical Projection and Occlusion Deduction}\label{sec:cylindrical projection} We apply the non-linear cylindrical projection to encode LiDAR points from 3D continuous space on a 2D discretized plane. It finds the corresponding 2D coordinates of each point in the 3D cloud and allows the processing of the 3D point cloud in a 2D form with the chosen feature channel. Given a point $\mathbf{p}=[x,y,z]^\top$ in a 3D point cloud, the corresponding pixel on panorama is given by \begin{equation}
\left[\begin{array}{c}
i \\
j
\end{array}\right]
=
\left[\begin{array}{c}
\frac{h}{\delta_h}  \arctan \frac{y}{x}\vspace{1ex} \\
\frac{h}{\delta_v}  \arctan \frac{z}{\sqrt{x^{2}+y^{2}}}
\end{array}\right]
\label{eq:cylindrical_projection}
\end{equation} where the vector $\bm v = [i,j]^\top$ corresponds to the 2D coordinates on the panorama and $h$ is the scaling parameter, determining the size of the image. The horizontal and vertical angular resolutions $\delta_h$ and $\delta_v$ are acquired from the sampling frequency of the LiDAR. Once calculated, it serves as a look-up table. Any input 3D points can find the corresponding counterpart on the panorama.

\begin{figure}[t]
\centering
\subfigure[The camera image.]{
\includegraphics[width=0.33\textwidth]{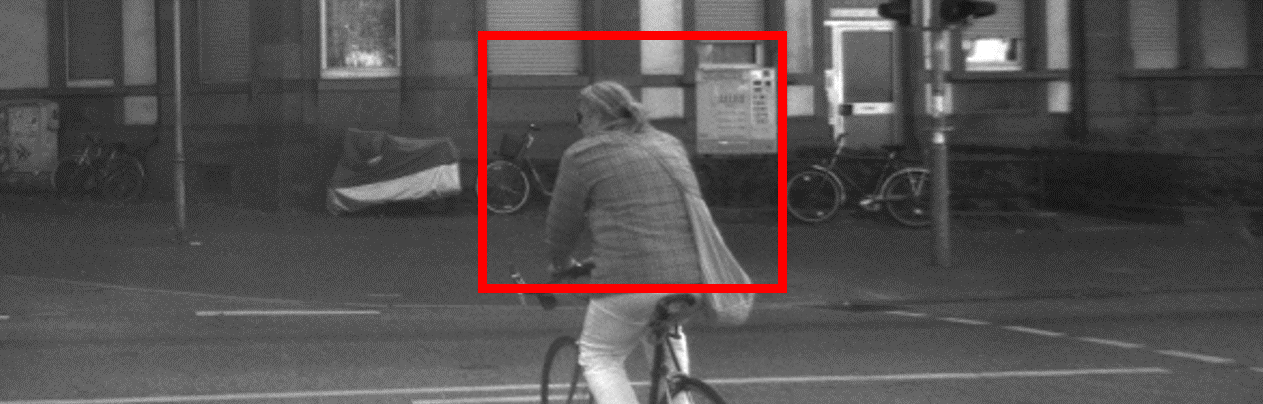}}

\subfigure[{\scriptsize w/o occ. deduction}]{
\includegraphics[width=0.16\textwidth]{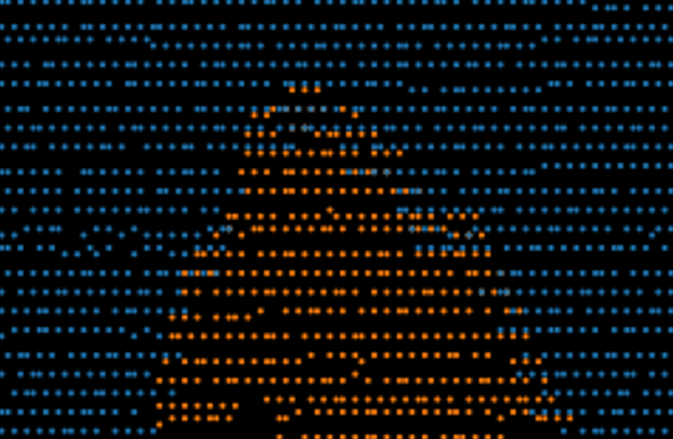}}
\subfigure[{\scriptsize Canny Edge}]{
\includegraphics[width=0.13\textwidth]{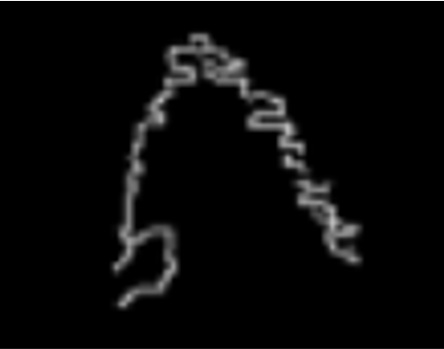}}

\subfigure[ w/ occ. deduction]{
\includegraphics[width=0.16\textwidth]{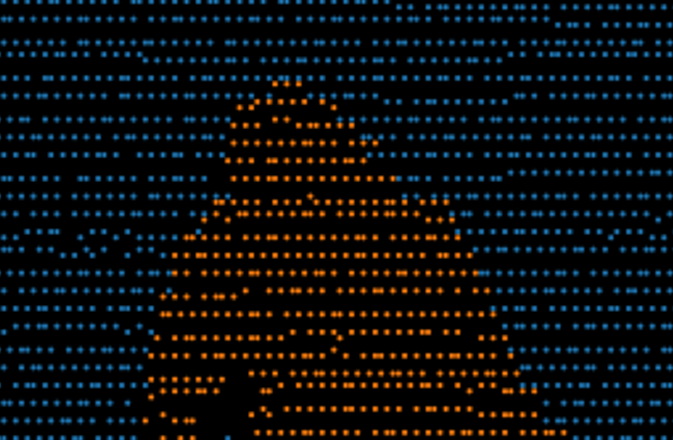}}
\subfigure[{\scriptsize  Canny Edge}]{
\includegraphics[width=0.13\textwidth]{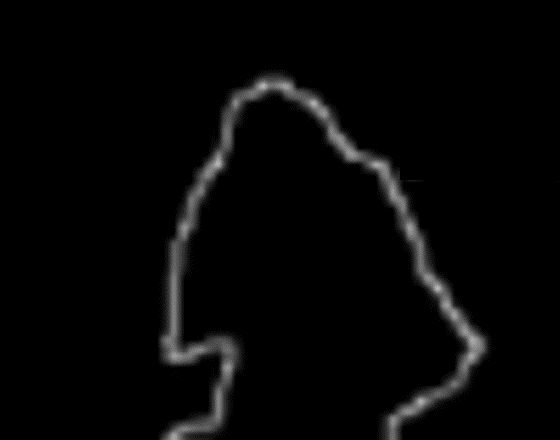}}
\caption{\small An example of the occlusion effect and its influence on the edges. The red box region in (a) shows a cyclist in the foreground, against the background wall, and (b) and (c) show the sparse image of the cyclist (without occlusion deduction) and the corresponding edge map respectively. The effect of the proposed occlusion deduction is shown in (d) and (e).}
\label{fig:occ_2}
\end{figure}

The foreground mask $D$ is calculated by a morphological closing operator to highlight the foreground regions. We denote the panorama only with foreground pixels as $\mathbf{u}_{obj}$. All the background points labeled as background while on the foreground mask are discarded from the original point cloud $\bm \psi$. After the morphological operation, the resulting map is improved on the edges, as shown in Fig \ref{fig:occ_2}. Algorithm \ref{algo:cp_occlusion} shows the whole point cloud preprocessing.

\begin{algorithm}
\SetAlgoLined
 \KwIn{Point cloud $\bm{\psi}$}
 \KwOut{The filtered sparse panorama $\mathbf{u}_L$}
 
 $\bm \psi_1 \leftarrow$ RANSAC plane segmentation of $\bm \psi$. \\ 
 
 $\bm \psi_2 \leftarrow$ DBSCAN clustering of $\bm \psi_1$. \\
 Calculate average depth $d_k$ of $k$ clusters $\mathcal{C}_k$ in $\bm \psi_2$. \\
 Assign object labels (\ref{eq:obj_values}).\\
  $\mathbf{u}(i, j) \leftarrow$ Cylindrical projection of $\bm \psi$ (\ref{eq:cylindrical_projection}). \\ 
 $\mathbf{u}_{obj} \leftarrow \{\mathbf{u}_L(i, j) \mid o = 2\}$. \\ 
 Object mask $\bm D \leftarrow$ morphological closing of $\mathbf{u}_{obj}$.\\
 \For{$\mathbf{p}(i,j) \in \bm \psi$}{
 \If{$\mathbf{p}(i,j) \in \mathcal{C}_k$, $k \neq 2$ and 
 $\bm D(i,j)=2$}
 {$\mathbf{u}_L(i, j)=0$.
 }}
 \caption{Point cloud preprocessing algorithm}
 \label{algo:cp_occlusion}
\end{algorithm}


\subsection{Dense Map Completion} \label{sec:dense_map_comp}


\begin{figure}[t]
\centering
\includegraphics[width=0.38\textwidth]{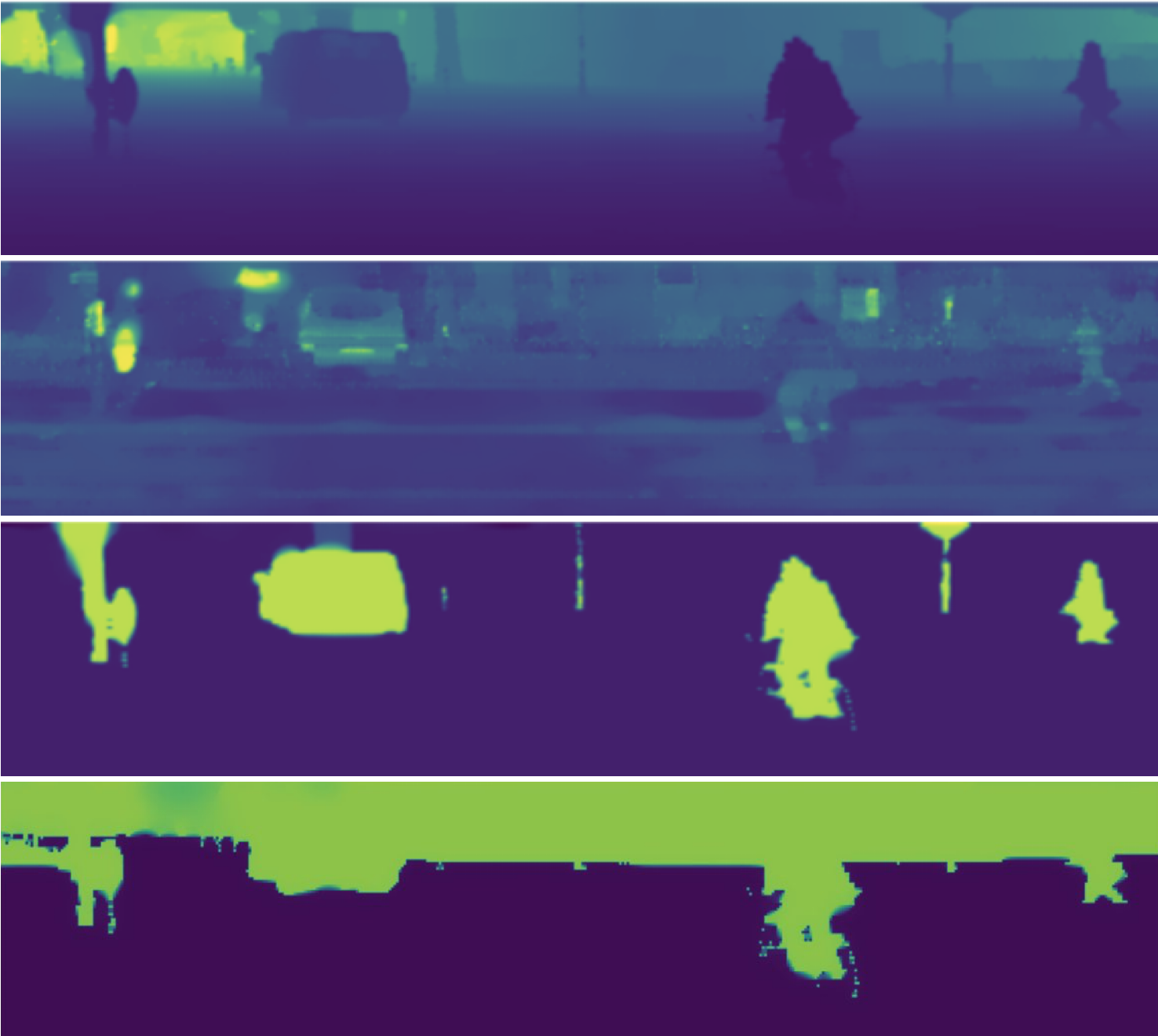}
\caption{The dense maps of depth, reflectivity, near-field objects and ground plane.} 
\label{fig:dense_maps}
\end{figure}
The LiDAR information is sparse and cluttered. Projecting all the 3D points on the cylindrical plane only solves the latter problem, by encoding them as images. Construction of dense maps solves the sparsity problem so that the LiDAR information can be analyzed as ordinary images. Such recovery can be formulated as image or signal upsampling. In Fig \ref{fig:dense_maps}, we give examples of the dense maps of four feature channels. The sparse depth map, reflectivity map, foreground map and ground plane map are denoted as $\mathbf{u}_d$, $\mathbf{u}_r$, $\mathbf{u}_o$ and $\mathbf{u}_p$, respectively. Suppose $\bm \phi \in \mathbb{R}^{N_x \times N_y}$ is the desired dense panoramic image, where $N_x$ and $N_y$ indicate the dimensions of the map along the $x$ and $y$ dimensions,  $\mathbf{u}_L$ represents the 2D sparse panoramic image obtained from the 3D point cloud, and $\mathbf{H}$ the known binary masking matrix, which is created based on the locations on the panorama with LiDAR projection. We then have 

\begin{equation}
\begin{aligned}
\hat{\bm{\phi}} = & \operatorname*{arg\,min}_{\bm{\phi} \in \Phi} \; \Big\{  \left\|\mathbf{u}_L - \mathbf{H} \odot \bm{\phi}\right\|_2^2 \\
& + \lambda \sum_{i=1}^{N_x} \sum_{j=1}^{N_y} 
\Big( \left\|[\boldsymbol{\nabla}_x \bm{\phi}]_{i,j}\right\|_1 + \left\|[\boldsymbol{\nabla}_y \bm{\phi}]_{i,j}\right\|_1 \Big)
\Big\}
\end{aligned}
\label{eq:dense_map_completion}
\end{equation} where $\odot$ is the element-wise matrix multiplication. We formulate the data fidelity within an $\ell_2$ norm that constrains the feature values in the given positions by the masking matrix. The second term accumulates the total gradient in both $x$ and $y$ directions of the desired map. By minimizing the redundant gradient changes due to noise or sparse information, we obtain smoothness in the whole image, while the sharp edges are kept since the $\ell_1$ norms promote sparsity. The complete problem is formulated as a variant of the ROF (Rudin-Osher-Fatemi) image restoration model \cite{TV}. The dense panoramic map can be recovered by minimizing the data fidelity loss and total variation norms, which can be solved in very few iterations \cite{FISTA} \cite{split} \cite{EPFL}.

\subsection{Point Cloud Edge Extraction}\label{sec:pc_edge}




\begin{figure}[t]
\centering
\includegraphics[width=0.38\textwidth]{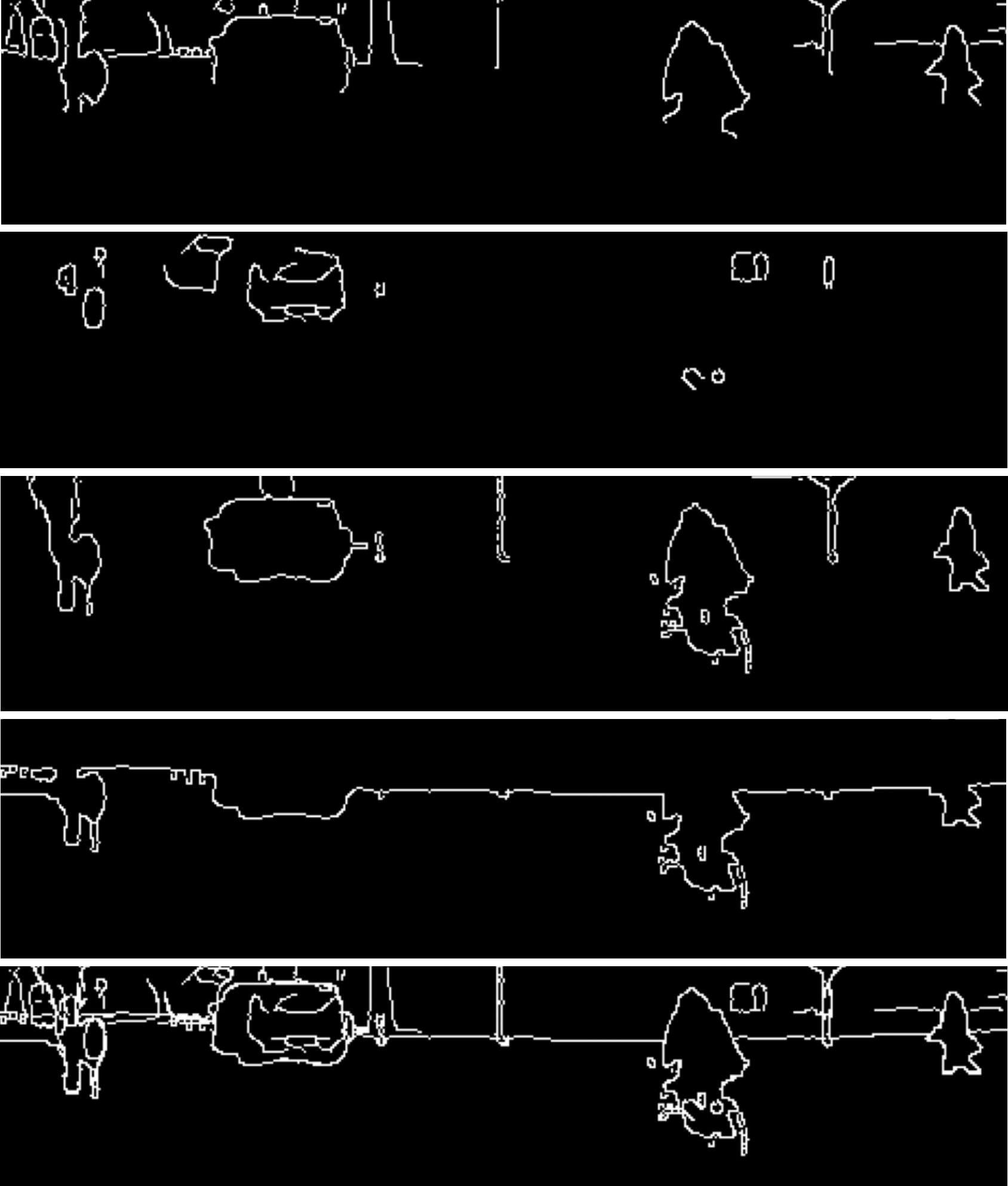}
\caption{The edge maps of depth, reflectivity, foreground objects, ground plane and the mixture of four, respectively. The depth edges provide the depth discontinuity of the environment. The reflectivity edges add details to the patterns of the objects. The object edges provide full contour of the foreground objects and the ground provides a clear horizontal boundary. The mixture contains the high-quality edges solely from LiDAR points.}
\label{fig:edge_mix}
\end{figure}

We use Canny edge detection \cite{canny} for multi-feature map processing. The output edge maps are $E_d$, $E_r$, $E_o$, and $E_p$, denoting the depth, intensity, object, and ground edge, respectively.
The mixed edge map $E_L$ is the fused map of the equally weighted edge maps after the Canny edge detection, shown in (\ref{eq:mix_edge}).
\begin{equation}
    \bm{E}_L(i,j) = \frac{1}{4}\sum_{k \in\{d, r, o, p\}} \bm{E}_{k}(i, j),
    \label{eq:mix_edge}
\end{equation} where the left-hand side denotes the mapping from a given point to the intensity on the edge map. The mixed map provides more details, bringing rich edge information for further alignment, as shown in Figure \ref{fig:edge_mix}. Therefore, any given 3D point $\bm p = (x, y, z)$ can be encoded via cylindrical projection to a 2D pixel $\bm v = (i, j)$ on the panoramic image, with edge intensity $\bm E_L(i, j)$ or $\bm E_L(\bm p)$.




\subsection{Objective Function and Optimization} \label{sec:opt} Assume there are $N$ points in the point cloud. Among them, $N_e$ points are edge points selected with non-zero intensities. If an edge point is projected onto the intensity edge of the camera image, the counter $N_m$, which stands for the number of matched points, will accumulate. This ratio is defined as $\frac{N_m}{N_e}$ and interpreted as precision in a confusion matrix. Ideally, the correct calibration parameters will result in the highest ratio of $\frac{N_m}{N_e}$. We design the cost function by constructing the linear correlation between edge intensities of camera images and edge intensities acquired from multi-feature LiDAR maps, i.e., 
\begin{equation}
J(\bm \theta)= \frac{N_m}{N_e} \sum_{n=1}^{N} \bm E_C(\bm u_n) \cdot E_L(\bm p_n).
\label{eq:cost_own}
\end{equation}
where $\bm u_n$ is the projected pixel of a 3D point given the projection matrix $T_{\bm \theta}$ derived from $\bm \theta$, according to (\ref{eq:extrinsic}). To optimize such a non-concave function, we adopt the conventional gradient ascent method.

The non-trivial part of using the gradient ascent method is that the gradient is hard to be derived analytically since the statistical counts and indirect projection features cannot be straightly related to the calibration parameters in the form of functions. Therefore, one feasible approach is to use the numeric gradient instead, 
\begin{equation}
\mathbf{G}=\nabla J(\bm{\theta})=\frac{J(\bm{\theta}+\Delta \mathbf{h})-J(\bm{\theta}-\Delta \mathbf{h})}{2 \cdot \Delta \mathbf{h}}, 
\label{eq:gradient_value_own}
\end{equation}
and the numerical derivatives can be calculated for a small $\Delta \mathbf{h}$. 




\section{Experiments}\label{sec:experiments}

We now evaluate the performance of the proposed Multi-FEAT pipeline based on the open-source KITTI dataset \cite{KITTI}, a benchmark in autonomous driving. 
We compare the objective function of our approach qualitatively with several existing targetless solutions, Pandey et al. \cite{MI}, Levinson et al. \cite{Levinson} and Castorena et al. \cite{castorena}, respectively. The shapes of objective functions with rotation and translation displacements are presented. We use the interval $-0.3$ to $0.3$ rads for rotational parameters and for the translation parameters, $-0.3$ to $0.3$ meters. Note that the objective function in Castorena et al. needs minimization, while the rests need maximization. For clarity, we normalize the cost functions into intervals $[0, 1]$. We use a threshold $\gamma=0.12$ for RANSAC, the minimum number of points to form cluster $m=8$, a range $R=0.25$ to form a cluster using DBSCAN and $\lambda=0.05$ in TV norm (\ref{eq:dense_map_completion}). 






\subsection{Single-frame evaluation} To analyze the effectiveness of our proposed Multi-FEAT pipeline, we select unique urban scenarios within the scope of the KITTI dataset. 

\begin{figure}[t]
\centering
\includegraphics[width=0.45\textwidth]{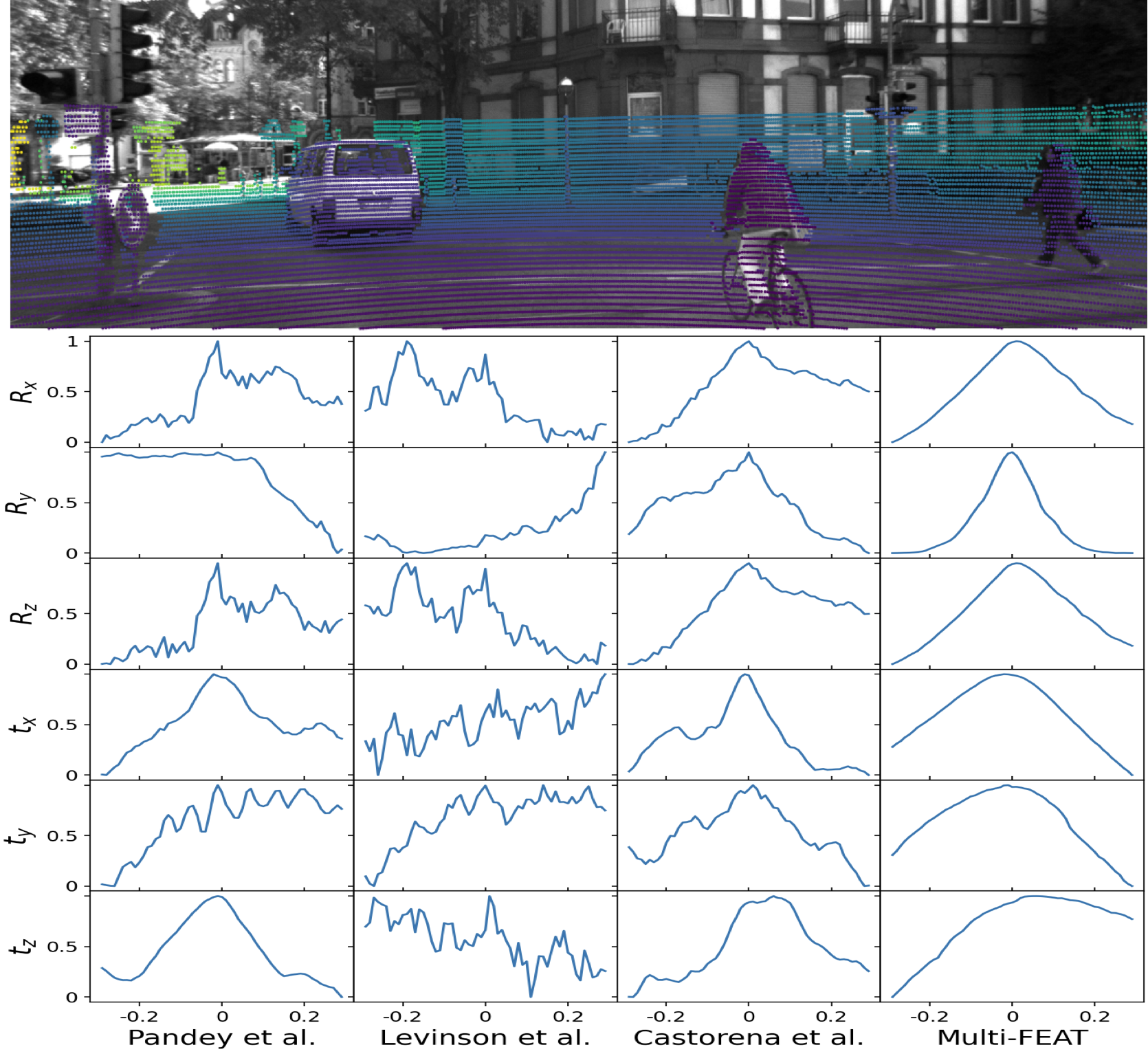}
\caption{\textit{Scenario 1 (Urban Road)} shows a typical example of a road junction, where there are pedestrians, vehicles, cyclists and buildings in the backdrop, all within the field of view of the experimental vehicle.}
\label{fig:curves_comparison_05}
\end{figure}

\subsubsection{Scenario 1: Urban Road} In this scenario, shown in Fig. \ref{fig:curves_comparison_05}, pedestrians are crossing the road, the vehicle in the middle of FOV with the cyclist on the road, and the tram rail, the traffic signs, and buildings in the background. We notice in Fig. \ref{fig:curves_comparison_05}, that Pandey et al.'s method in the first column provides the local maxima around the ground truth values for almost all six parameters. However, it is generally non-smooth. The global optimum is distinct in the method by Levinson et al.. However, various local maxima would impede the gradient-based optimizers from achieving the actual value. Castorena, et al.'s method also has some limitations since it overlooks the occlusion effect, and it is evident that the shape of the cost function is not smooth. Compared to the other methods, the proposed Multi-FEAT shows a smoother curve for almost all the parameters, with a peak almost centering at zero.

\begin{figure}[t]
\centering
\includegraphics[width=0.44\textwidth]{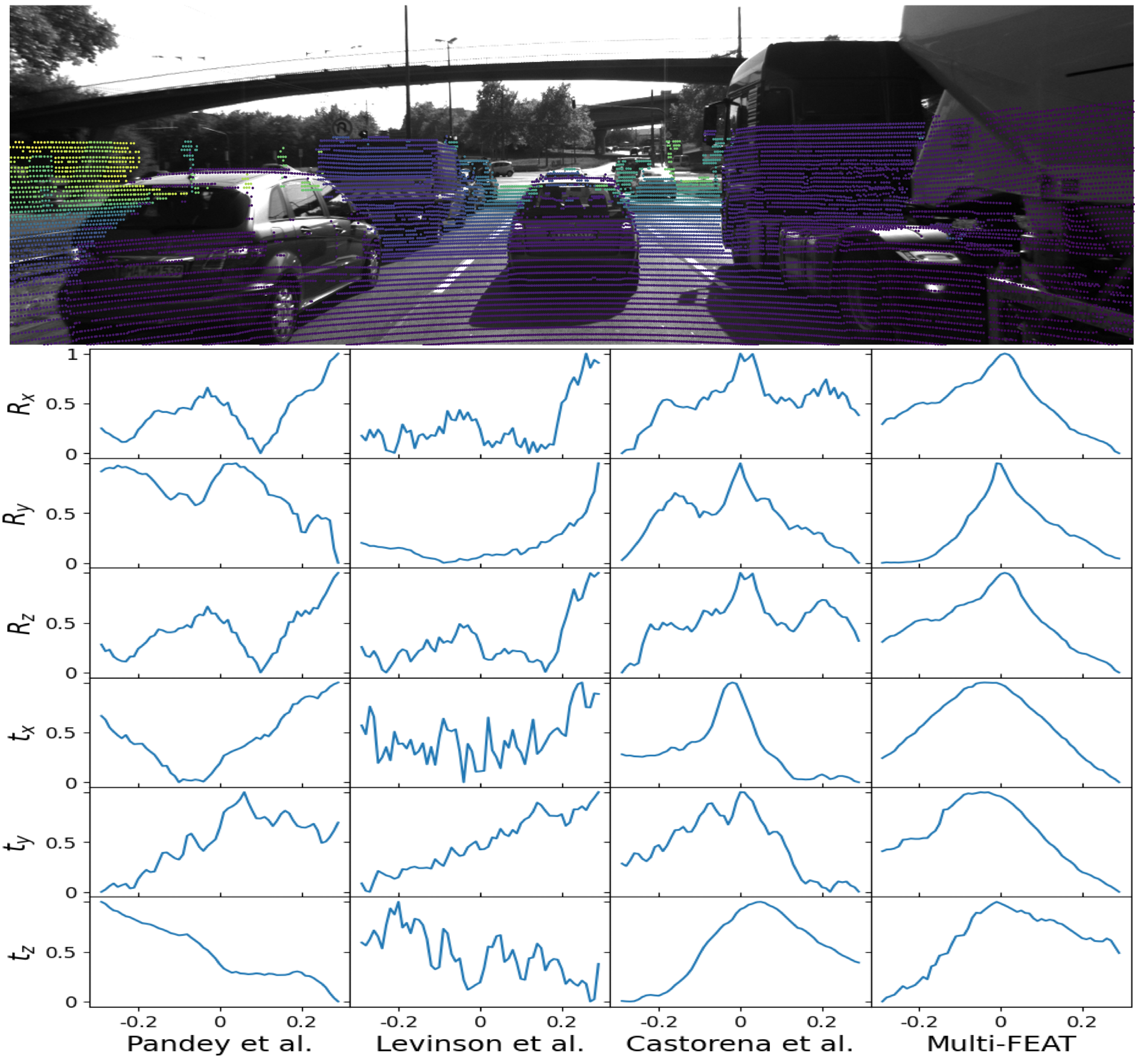}
\caption{\textit{Scenario 2 (Highway)} presents an example of highway scenario. LiDAR cannot see through the near-field vehicles around, therefore providing little information from farther objects.}
\label{fig:curves_comparison_52}
\end{figure}

\subsubsection{Scenario 2: Highway}
The second example is from the highway with vehicles in the surrounding. The vehicle on the left is too near to the onboard LiDAR. Hence many laser beams that should illuminate the vehicle's body are reflected elsewhere that the LiDAR cannot receive. The densely packed vehicles in the periphery make clustering and separating objects challenging. In Fig. \ref{fig:curves_comparison_52}, we show the image with laser points and the results. In these complex scenarios with closely spaced objects under a heavy shadow, the performance of Pandey et al.'s method is no longer robust. For $t_x$ and $t_z$, the mutual information assumption does not hold in this scenario. We observe numerous local optima in the case of the methods proposed by Levinson et al. and Castorena et al.. In contrast, our proposed approach shows a smooth objective function for estimating the rotation parameters. 

\subsubsection{Scenario 3: Neighbourhood alley} 
\begin{figure}[t]
\centering
\includegraphics[width=0.45\textwidth]{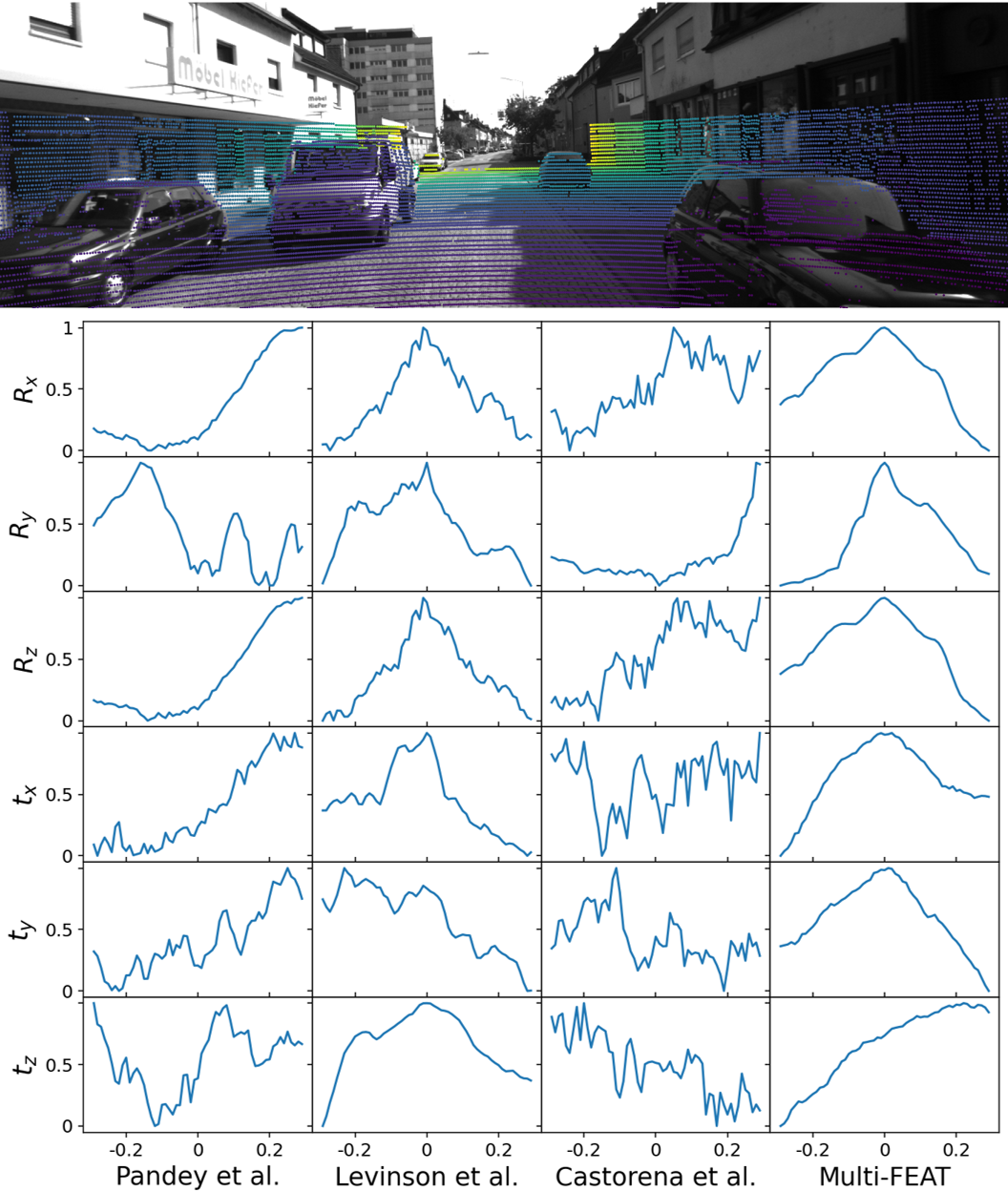}
\caption{\textit{Scenario 3 (Neighbourhood alley):} shows a small alleyway with cars parked alongside the road. Observe the stark change of brightness in the middle of the image, due to the shadow cast by the building on the right.}
\label{fig:curves_comparison_48}
\end{figure}

The third example is a narrow straight alley with parked cars along the road. In this case, there are no lanes on the road or other traffic signs. The shadow of the building on the right leaves a sudden transition of brightness in the middle of the image. Due to the electroplated surface of the vehicles, we observe a specular reflection on the cars in the near-field. Therefore a large number of LiDAR points are missing, and thus challenging the edge detection operation. The image and the results can be seen in Fig \ref{fig:curves_comparison_48}.

For all the parameters in the first column (Pandey et al.,), we observe a non-smooth objective function, which we suspect is due to the shadow in the image, which heavily influences the mutual information-based approach since the intensity-reflectivity pairs are no longer the same inside and outside shaded areas. The loss of object LiDAR points due to the specular reflection on the surface of near-field vehicles also degrades the results of all three prevalent edge alignment methods, but the proposed Multi-FEAT approach appears more favourable under these circumstances. Nevertheless, Multi-FEAT may lead to a biased estimate of the final results of the estimation of calibration parameters since the shape of curves, especially on $t_z$.

\subsubsection{Scenario 4: Suburban Area} 
\begin{figure}[t]
\centering
\includegraphics[width=0.45\textwidth]{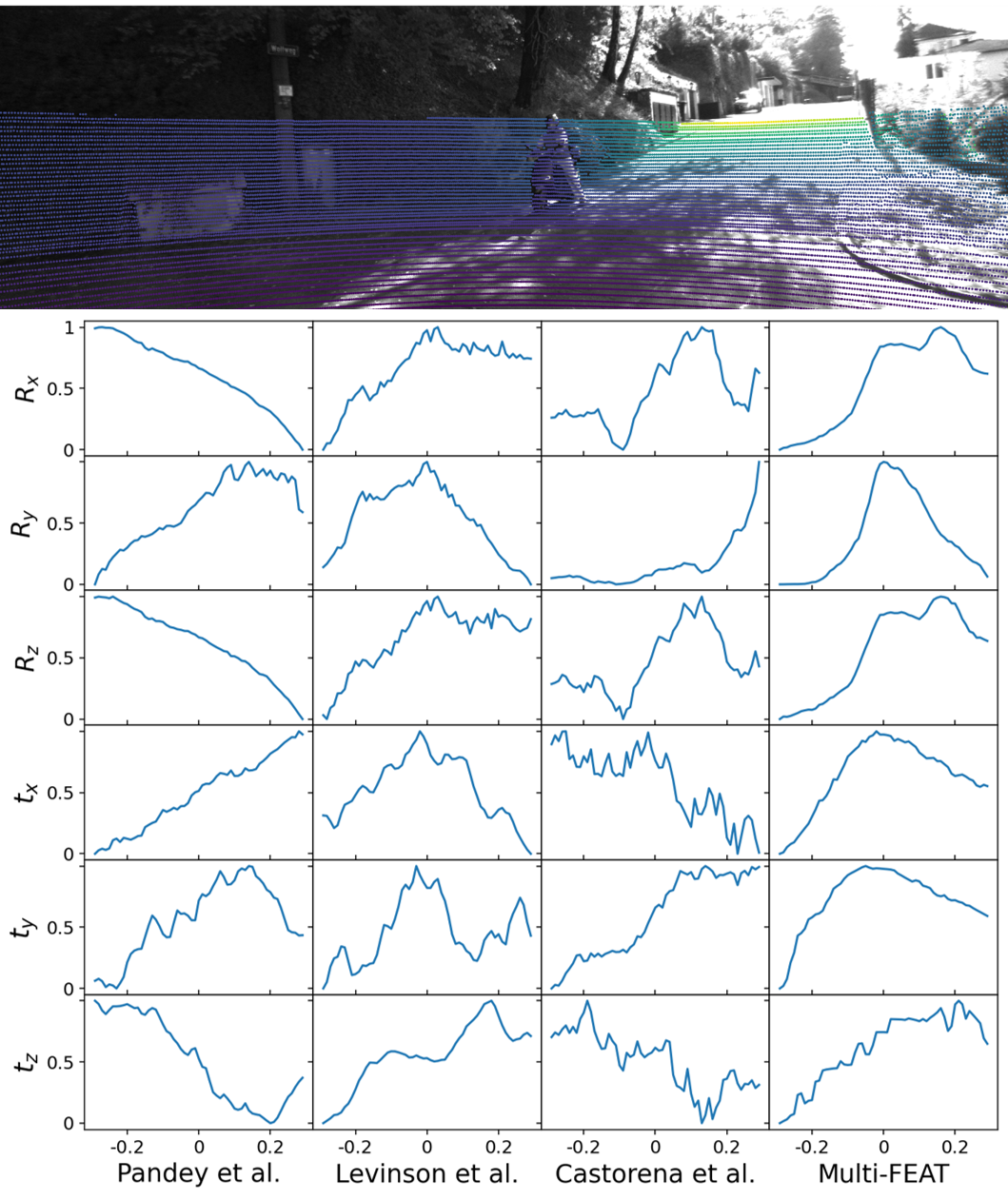}
\caption{\textit{Scenario 4 (Suburban landscape)} shows a complex environment for camera-LiDAR calibration algorithms. There are only few objects which provide valid edges for the calibration process, and the shadow from the trees negatively affects more than half of the camera image. There are also vines along the left wall, providing complex edges on the intensity image while almost invisible on the point cloud.}
\label{fig:curves_comparison_86}
\end{figure}

The forth experiment covers a complex environment in a suburban area. The experiment vehicle, with the on-board camera and LiDAR, faces a ramp with a positive inclination. Given this view, observe that the plane segmentation cannot extract both the flat ground and the ramp. Furthermore, some plane points participate in the object clustering, hence degrading the results. In the image, we notice the motorbike rider is coming down the ramp, and is completely hidden under the shade of the trees. In general, more than half of the image is corrupted by the shadow of the trees. A wire pole stands tall to the left of the image, but is close to the wall, with little depth discontinuity information to be exploited. The clustering fails to separate the motorbike rider and the wire pole effectively. Hence, in the FOV of the experimental vehicle, there are few valid objects for edge alignment. The results are illustrated in Fig \ref{fig:curves_comparison_86}.


Pandey et al.'s approach faces severe challenges since there is a weak correlation between image intensity and laser reflectivity, primarily due to the shadows. The other edge alignment methods need valid objects in the FOV, which is severely limited in this scenario. It is clear that none of the existing targetless methods offer a smooth objective function for any of the $6$ parameters. The performance of our proposed method is also severely affected in this complex environment, however the rotation parameter $R_y$ and translation parameters $\{t_x,t_y\}$ show promising results with a clear maximum. 



\begin{table}
\scriptsize
    \centering
    {\renewcommand{\arraystretch}{1.5}
    \begin{tabular}{c|c|c|c|c|c|c}
    \hline\hline
    \textbf{} & \textbf{$r_x$(rad)} & \textbf{$r_y$(rad)} & \textbf{$r_z$(rad)} & \textbf{$t_x$(m)} & \textbf{$t_y$(m)} & \textbf{$t_z$(m)}\\
    \hline
    \hline
    \text{Levinson et al.} & 0.018 & 0.026 & 0.017 & 0.036 &  0.074 & 0.054 \\
    \hline
    \text{Zhu et al.} & 0.011 & 0.007 & 0.010 & 0.048 &  0.033 & 0.036 \\
    \hline
    \text{Multi-FEAT} & 0.006 & 0.005 & 0.006 & 0.017 & 0.025 & 0.041 \\
    \hline\hline
    \end{tabular} }
    \caption{Results of the Multi-frame evaluation of different methods. The value in each block is the mean absolute error of the estimate.}
    \label{tab:results}
\end{table}

\subsection{Multi-frame evaluation}  In this section, we aggregate multiple frames for optimization. More concretely, we select $N = 30$ frames during which the $6$ parameters are constant and provided in the KITTI dataset. We compare our results with another recent edge alignment-based work with only car-labeled semantic edges \cite{semantic}. Since it adopted a self-trained PSP-Net \cite{PSP-Net}, we cannot precisely replicate the full workflow. Instead, we use the same semantic segmentation pipeline Grounded-SAM, with only the car semantic as the prompt. We also take depth discontinuity-based method \cite{Levinson} into the qualitative comparison.

Table \ref{tab:results} shows the quantitative 
results of three different edge alignment methods with different choices on feature selection. Multi-FEAT outperforms the other two listed methods. One possible reason could be that Multi-FEAT aggregates data from more features than the other two. For camera features, Levinson et al. use intensity edge, Zhu et al. apply car semantic edge, while we include both of them, with extensions on semantic types. As for LiDAR points, both the above mention methods exploit geometry features, while ours incorporate not just multi-layer geometric information, but also LiDAR reflectivity.

\begin{figure}[t]
\centering
\includegraphics[width=0.40\textwidth]{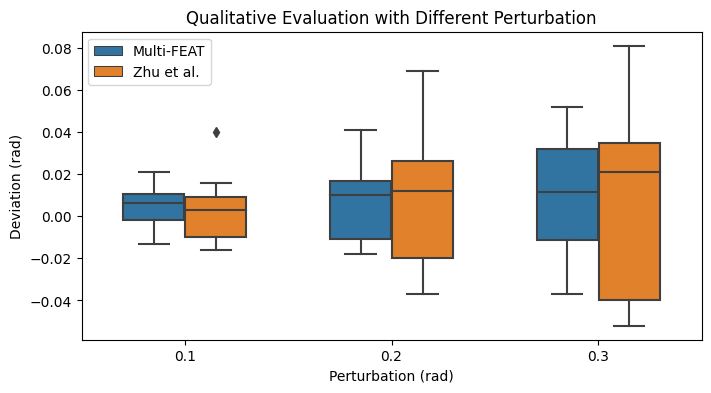}
\caption{The figure shows the results of the proposed method and Zhu et al.'s method \cite{semantic} in box plots with different angle perturbations.}
\label{fig:perturbation}
\end{figure}

To prove the adaptivity of our method, we test with different initial values of rotation angles, in 10 different scenes from KITTI, shown in Fig. \ref{fig:perturbation}. When the deviation is small, the estimation results have comparable biases, since the edges are almost aligned. However, when given larger deviations as initial guesses, Multi-FEAT tends to have less biases and smaller variance. Because when there are few cars in the scene, or when the cars are far from the sensors, the accuracy and the robustness of car-only assumption is challenged. Also, since \cite{semantic} uses the full non-ground point cloud for scoring, rather than the edges of the point cloud, it might face certain local optima when multiple cars jammed together, as the height mask covers the majority of the FOV.
Multi-FEAT, on the other hand, collects more information for edge alignment, therefore has better performance in different road scenarios.

\section{Conclusion}\label{sec:conclusion}
This paper proposes Multi-FEAT for targetless extrinsic calibration between the monocular camera and LiDAR. It encodes the LiDAR information into 2D by cylindrical projection, and tackles the occlusion effect with clustering and the morphological closing. By combining the depth, reflectivity, foreground objects and ground edge features and formulating a statistical model, we enriched the information of the LiDAR point cloud, thereby obtaining a better fit with designed objective function. We perform simulations to compare Multi-FEAT algorithm with several state-of-the-art targetless calibration methods using the KITTI dataset, emphasizing the advantages of our approach in diverse scenarios. 

Although the adopted image segmentation pipeline, Segment Anything, represents the state-of-the-art segmentation solution, the granularity of the network and object-level semantic classification are also worthy of investigation for camera-LiDAR calibration purposes. We also anticipate future studies focused on encoding LiDAR features to achieve better similarity compared to images. Furthermore, exploring the intrinsic bias of edge alignment calibration methods is worthwhile, especially when dealing with significant parallax between sensors or limited mutual FOVs.






\bibliographystyle{unsrt}
\bibliography{ref}

\end{document}